\documentclass[10pt]{article}
\usepackage{amsmath}
\usepackage{amssymb}
\usepackage{natbib}

\begin{document}
\bibliographystyle{plainnat}
\setcitestyle{numbers,square}

\title{\normalsize{GRAVITATIONAL QUANTIZATION OF SATELLITE ORBITS    \\
                   IN THE GIANT PLANETS}
                  }

\author{Vassilis S. Geroyannis \\
        Department of Physics, University of Patras, Greece          \\  
        vgeroyan@upatras.gr}

\maketitle

\begin{abstract}
A fundamental assumption in the so-called ``global polytropic model'' is hydrostatic equilibrium for a system of planets or statellites. By solving the Lane-Emden differential equation for such a system in the complex plane, we find polytropic spherical shells defined by succesive roots of the real part $\mathrm{Re}(\theta)$ of the Lane-Emden function $\theta$. These shells seem to be appropriate places for accomodating planets or satellites. In the present study, we apply the global polytropic model to the systems of satellites of the giant planets: Jupiter, Saturn, Uranus, and Neptune.  
\end{abstract}

\section{Introduction}
\label{intro}
We study the issue of the gravitational quantization of orbits in the systems of satellites of the giant planets by using classical mechanics. As discussed in a recent investigation (\citep{GVD14}, Secs.~1 and 6), these systems are treated within the framework of the ``global polytropic model'', assumed to obey the equations of hydrostatic equilibrium. 
These equations yield the well-known Lane--Emden equation, which is solved in the complex plane by applying the so-called ``complex plane strategy'' (readers interested in this issue can find full details in \citep{GKAR14}), and gives as solution the complex Lane--Emden function $\theta$.
There is in fact only one parameter to be adjusted for a particular polytropic configuration defined by $\theta$: the polytropic index $n$ of the central body (star or planet). A general algorithm for computing an optimum value, $n_\mathrm{opt}$, of this polytropic index is given in Sec.~\ref{gpm}. 

Alternative studies concerning quantized orbits of planets and satellites can be found in the following investigations. In \citep{CK08}, the authors solve analytically for the equilibrium structure of the midplane of a gaseous isothermal disk, incorporating in the Lane--Emden equation the effects of self-gravity, differential rotation, and thermal pressure. Then, they adopt a four-parameter analytic solution as ``baseline'' and use the rotation profile of the baseline in order to compute the ``oscillatory equilibrium solution'' obeying the physical boundary conditions at the center. They achieve fitting the density maxima of the solution to the planetary orbits in the present solar system. This model is used in \citep{CK08b} for studying the planetary system of the star 55 Cnc.
Regarding the Titius-Bode law, or modifications of this law, interested readers can find further issues in \citep{GD94}, \citep{HT97}, \citep{BL13}, and \citep{HB14}. Several methods and theories on quantized orbits are given in \citep{PPL08}, \citep{G07}, \citep{AF97}, \citep{RR98}, \citep{OMC04}, and \citep{HSG98} (a brief review on such considerations is given in \citep{GVD14}, Secs.~1 and 6). Comparisons of predictions of planetary quantization by some existing methods can be found in \citep{C04}.  

\section{Global Polytropic Model}
\label{gpm}
A detailed review on the complex-plane strategy and the complex Lane--Emden function $\theta$ can be found in \citep{GKAR14} (Sec.~3.1; see also \citep{GVD14}, Sec.~2); preliminaries regarding the global polytropic model can be found in \citep{GVD14} (Sec.~3). For convenience, we use here the same definitions and symbols with those adopted in \cite{GVD14}. 

The real part $\bar{\theta}(\xi)$ of the complex function $\theta(\xi)$ has a first root at $\xi_1 = \bar{\xi_1} + i \, \breve{\xi_0}$, a second root at $\xi_2 = \bar{\xi_2} + i \, \breve{\xi_0}$ with $\bar{\xi_2} > \bar{\xi_1}$, a third root at $\xi_3 = \bar{\xi_3} + i \, \breve{\xi_0}$ with $\bar{\xi_3} > \bar{\xi_2}$, etc. The polytropic sphere of polytropic index $n$ and radius $\bar{\xi}_1$ is the central component of a resultant polytropic configuration, of which further components are the polytropic spherical shells $S_2$, $S_3$, \dots, defined by the pairs of radii $(\bar{\xi}_1, \, \bar{\xi}_2)$, $(\bar{\xi}_2, \, \bar{\xi}_3)$, \dots, respectively.
Each polytropic shell can be considered as an appropriate place for a planet or satellite to be ``born'' and ``live''; and the most appropriate ``accomodation distance'' $\alpha_j \in [\bar{\xi}_{j-1},\,\bar{\xi_j}]$ is that at which $|\bar{\theta}|$ takes its maximum value inside $S_j$, 
\begin{equation}
\mathrm{max}|\bar{\theta}[S_j]| = |\bar{\theta}(\alpha_j + i \, \breve{\xi}_0)|.
\label{maxth} 
\end{equation}

An algorithm has been developed in \cite{GV94}, called A[n], for computing the optimum polytropic index $n_\mathrm{opt}$ for a star with a system of planets, or for a planet with a system of satellites. We present here a more general form of A[n], applied to $N_\mathrm{P}$ members $P_1,\,P_2,\,\dots,\,P_{N_\mathrm{P}}$ of a system with $N_\mathrm{P}$ prescribed distances $A_{1} < A_{2} < \dots < A_{{N_\mathrm{P}}}$ from the central body. To compute $n_\mathrm{opt}$, we work as follows. \\
\textbf{A[n]--1}.~For an array $\{ n_i \}$ of polytropic indices with elements 
\begin{equation}
n_i = n_1 + H_n \, (i-1), \qquad i = 1, \, 2, \, \dots, \, N_n,
\label{Nn}
\end{equation}
we compute the corresponding array of distances $\{\alpha_{j}(n_i)\}$, $j=2,\,3,\,\dots,\,L_n$, at which members of the system can be accomodated, with the integer $L_n$ taken sufficiently large. \\
\textbf{A[n]--2}.~For each array $\{\alpha_{j}(n_i)\}$, we compute the two-dimensional array of distance ratios
\begin{equation}
D(n_i;\,j,\,k) = \, \frac{\alpha_{j}(n_i)}{\alpha_{k}(n_i)}, \qquad 
j = 2,\,3,\,\dots,\,L_n, \qquad k = 2,\,3,\,\dots,\,L_n,
\label{Darray}
\end{equation} 
\textbf{A[n]--3}.~We scan the arrays $D(n_i;\,j,\,k)$ in order to find values $n_\ell$, $\ell=1,\,2,\,\dots,\,N_\ell$, generating ``proper levels'' related to the $N_\mathrm{P}-1$ ratios 
\begin{equation}
R_{1,2} = A_{1}/A_{2}, \, R_{1,3} = A_{1}/A_{3}, \, \dots, 
                       \, R_{1,N_\mathrm{P}} = A_{1}/A_{{N_\mathrm{P}}}.
\label{ratios}
\end{equation}
 By definition, $N_\mathrm{P}-1$ elements $D(n_\ell;\,q_1,\,q_2)$, $D(n_\ell;\,q_1,\,q_3)$, \dots, $D(n_\ell;\,q_1,\,q_{N_\mathrm{P}})$ --- where the $N_\mathrm{P}$ indices $q_1,\,q_2,\,\dots,\,q_{N_\mathrm{P}}$ obey the relation $q_1 < q_2 < \dots < q_{N_\mathrm{P}}$ --- constitute a proper level $\{n_\ell;\,q_1,\,q_2,\,\dots,\,q_{N_\mathrm{P}}\}$ if the following $N_\mathrm{P}-1$ conditions            
\begin{equation}
\Delta(D(n_\ell;\,q_1,\,q_2)) =
100 \times \frac{|R_{1,2} - D(n_\ell;\,q_1,\,q_2)|}{R_{1,2}} \, \leq \tau,
\end{equation}
\begin{equation}
\Delta(D(n_\ell;\,q_1,\,q_3)) =
100 \times \frac{|R_{1,3} - D(n_\ell;\,q_1,\,q_3)|}{R_{1,3}} \, \leq \tau,
\end{equation}
$\qquad \qquad \qquad \dots$
\begin{equation}
\Delta(D(n_\ell;\,q_1,\,q_{N_\mathrm{P}})) =
100 \times \frac{|R_{1,N_\mathrm{P}} - D(n_\ell;\,q_1,\,q_{N_\mathrm{P}})|}{R_{1,N_\mathrm{P}}} \, \leq \tau,
\end{equation}
are valid within a prescribed tolerance $\tau$. \\
\textbf{A[n]--4}.~For each proper level $\{n_\ell;\,q_1,\,q_2,\,\dots,\,q_{N_\mathrm{P}}\}$, we calculate the corresponding $N_\mathrm{P}$ absolute percent errors
\begin{equation}
\Delta(n_\ell;\,q_1) = 100 \times \frac{|A_{1} - \alpha(n_\ell;\,q_1)|}{A_{1}},
\end{equation}
\begin{equation}
\Delta(n_\ell;\,q_2) = 100 \times \frac{|A_{2} - \alpha(n_\ell;\,q_2)|}{A_{2}},
\end{equation}
$\qquad \qquad \qquad \qquad \quad \ \ \dots$
\begin{equation}
\Delta(n_\ell;\,q_{N_\mathrm{P}}) = 100 \times \frac{|A_{{N_\mathrm{P}}} - \alpha(n_\ell;\,q_{N_\mathrm{P}})|}{A_{{N_\mathrm{P}}}},
\end{equation}
and their sum,
\begin{equation}
\Delta(n_\ell;\,q_1,\,q_2,\,\dots,\,q_{N_\mathrm{P}}) = 
       \Delta(n_\ell;\,q_1) + \Delta(n_\ell;\,q_2) + \dots + 
       \Delta(n_\ell;\,q_{N_\mathrm{P}}).
\end{equation}
\textbf{A[n]--5}.~Among all proper levels $\{n_\ell;\,q_1,\,q_2,\,\dots,\,q_{N_\mathrm{P}}\}$, we localize the case with the ``minimum sum of absolute percent errors'', 
$\Delta_\mathrm{min} \bigl(n_L;\,q_1,\,q_2,\,\dots,\,q_{N_\mathrm{P}}\bigr)$,
and we identify the optimum polytropic index $n_\mathrm{opt}$ with that particular value $n_L$, i.e. $n_\mathrm{opt} = n_L$. 

\section{Polytropic Models Simulating Giant Planets}  
As discussed in \citep{CW09} (p.~65; see also \citep{CH14}, Sec.~1), a value $n \sim 2.5$ of the polytropic index is appropriate for molecular hydrogen and, accordingly, for the early stages of Jupiter's contraction. In \citep{GV94} (Sec.~2; see also \citep{GVD14}, Sec.~5.1), the algorithm A[n] has been driven to scan the array $\{n_i\}$ with elements
\begin{equation}
n_i = 2.40 + 0.01 \, (i-1), \qquad i = 1, \, 2, \, \dots, \, 21 
\label{Nn-pre}
\end{equation}
for finding $n_\mathrm{opt}$. Due to that prescribed interval of $n$ values, A[n] has determined a ``local optimum'', $n_\mathrm{opt}(\mathrm{Jupiter}) = 2.45 \in [2.40, \, 2.60]$, in the sense that the value $\Delta \bigl( n_L=2.45;\,q_1,\,q_2,\,\dots,\,q_{N_\mathrm{P}} \bigr)$ represents a minimum among the elements of the array $\{n_i\}$. 
During their evolution, however, the giant planets have developed several layers. In the bibliography (see e.g. \citep{HH83a} for two- and three-layer models; \citep{HH83b} and \citep{HM89} for interior models of the giant planets; \cite{CH14} for composite polytropic models appropriate to study the giant planets), the layers mostly discussed are: (i) a core consisting of iron and rocks, (ii) a surrounding mantle mainly consisting of metallic hydrogen and helium, and (iii) an outer envelope mainly consisting of molecular hydrogen and helium. Due to that particular evolution, appropriate values of $n$ for simulating the giant planets seem to be about $n \sim 1$ (\citep{Hor04}, Sec.~6.1.7 and references therein; \cite{HSD75}, Sec.~III; \citep{O62}, Sec.~26). 

In this study, we apply the general algorithm A[n], as described in Sec.~\ref{gpm}, to an array $\{n_i\}$ with elements
\begin{equation}
n_i = 0.700 + 0.001 \, (i-1), \qquad i = 1, \, 2, \, \dots, \, 1001.
\label{Nn-now}
\end{equation} 
So, we expect to find optimum values of the polytropic index for the giant planets in the interval 
\begin{equation}
\mathbb{I}_n = [0.700, \, 1.700]. 
\label{In}
\end{equation}

\section{The Computations}
\label{computations}
Preliminary details regarding computational environment (Fortran compiler, mathematical libraries used, etc.) are given in \citep{GVD14} (Sec.~4).

To solve the complex IVPs involved in this investigation, we use the code \texttt{DCRKF54} \cite{GV12}. This is a Runge--Kutta--Fehlberg code of fourth and fifth order modified for the purpose of solving complex IVPs, which are allowed to have high complexity in the definition of their ODEs, along contours (not necessarily simple and/or closed) prescribed as continuous chains of straight-line segments. Details on assigning values to input arguments of \texttt{DCRKF54} and on its usage are given in \citep{GVD14} (Sec.~4).

In this study, integrations proceed along the contour  
\begin{equation}
\mathfrak{C}_{\mathrm{JSUN}} = \{\xi_0 = (10^{-4},\,10^{-4}) \rightarrow 
                               \xi_\mathrm{end} = (8.0 \times 10^2,\,10^{-4})
                               \}
\label{CJSUN}
\end{equation}
for the satellite systems of the four giant planets. 
Apparently, this contour is of the special form~(8) of \citep{GVD14} (details on various contours and their characteristics are given in \cite{GV12}, Sec.~5).

\section{Numerical Results and Discussion}
\label{results}
Since physical interest focuses on real parts of complex quantities and functions, we will hereafter quote only such values and, for simplicity, we will drop overbars denoting real parts.

In this study, we first resolve the 1001 polytropic models counted in Eq.~(\ref{Nn-now}). Each model corresponds to a complex IVP defined by Eqs.~(5), (8), and (10) of \citep{GVD14}. As said in Sec.~\ref{computations}, all these IVPs are integrated by the Fortran code \texttt{DCRKF54}. We then apply the algorithm A[n] to all resolved models for the cases of the four giant planets of our solar system. 

Numerical results regarding optimum polytropic indices and respective proper levels are given in Tables~\ref{jupiter}--\ref{neptune}. In detail, Table~\ref{jupiter} shows results corresponding to the Jupiter's optimum case, with minimum sum of absolute percent errors 
\begin{equation}
\Delta_\mathrm{min} 
\biggl( n_\mathrm{opt}(\mathrm{J})=0.893;\,q_\mathrm{I}=7, \, q_\mathrm{E}=11, \,
                                               q_\mathrm{G}=18, \, q_\mathrm{C}=32 
      \biggr) \simeq 2.0
\label{DminJnow}
\end{equation}
(we use here the symbols of Table~\ref{jupiter}). Apparently, in the case of Jupiter the ratios $R_{i,j}$ of Eq.~(\ref{ratios}) are those corresponding to the distances of the Galilean satellites; namely,  
\begin{equation}
R_{1,2} = A_\mathrm{I}/A_\mathrm{E}, \ \ R_{1,3} = A_\mathrm{I}/A_\mathrm{G}, \ \ 
                                         R_{1,4} = A_\mathrm{I}/A_\mathrm{C}.
\label{ratiosJ}
\end{equation}
The smaller error is that for Io's distance, $\simeq 0.03\%$, and the larger one that for Europa's distance, $\simeq 0.9\%$. The error for the distance of the most massive satellite, i.e. Ganymede, is $\simeq 0.9\%$. The average error for the computed distances of the  Galilean satellites is $\simeq 0.5\%$.

On the other hand, the results of Table~IV in \citep{GV94} give for the (local) optimum case, when $n \in [2.40,\,2.60]$,    
\begin{equation}
\Delta \biggl( n=2.45;\,q_\mathrm{I}=3,\,q_\mathrm{E}=4,\,
               q_\mathrm{G}=5,\,q_\mathrm{C}=7 \biggr) \simeq 66.2,
\label{DminJpre}
\end{equation}
It should be stressed that this (local) minimum sum of absolute percent errors is $\sim 30$ times larger than the one quoted in Eq.~\eqref{DminJnow}. The large discrepancies found in \citep{GV94} can be interpreted as radii of satellite orbits holding for a ``proto-Jupiter'' with molecular hydrogen totally prevailing in its composition.

Regarding the next three giant planets, we implement the algorithm A[n] to tetrads of satellites, in order to test the ability of A[n] on ``predicting distances'' of other satellites, which are not included in the initial tetrad(s). In the case of Saturn we work with the tetrad: Dione, Rhea, Titan, and Iapetus. Thus the prescribed ratios are (using the symbols of Table~\ref{saturn})  
\begin{equation}
R_{1,2} = A_\mathrm{D}/A_\mathrm{R}, \ \ R_{1,3} = A_\mathrm{D}/A_\mathrm{T}, \ \ 
                                         R_{1,4} = A_\mathrm{D}/A_\mathrm{I}.
\label{ratiosS}
\end{equation}
The Saturn's optimum case has minimum sum of absolute percent errors
\begin{equation}
\Delta_\mathrm{min} \biggl( n_\mathrm{opt}(\mathrm{S}) = 1.239; \, q_\mathrm{D} = 6, \, 
         q_\mathrm{R} = 8, \, q_\mathrm{T} = 16, \, q_\mathrm{I} = 41 \biggr) \simeq 2.5.
\label{DminS}
\end{equation}
The smaller error is that for Rhea's distance, $\simeq 0.05\%$, and the larger one that for Titan's distance, $\simeq 1.8\%$; this satellite is also the most massive one in the system of Saturn. The average error in the computed distances of the four satellites is $\simeq 0.6\%$
 
Concerning the satellites Enceladus, Tethys, and Hyperion, which are not included in the initial tetrad, we remark that the distance of Enceladus is computed with a discrepancy $\simeq 4.5\%$, the distance of Tethys with a discrepancy $\simeq 1.5\%$, and that of Hyperion with a discrepancy $\simeq 0.9\%$. The total sum of percent errors, $\Delta_\mathrm{tot}$, for the seven satellites of Saturn considered here becomes 
\begin{equation}
\Delta_\mathrm{tot}(\mathrm{S}) \simeq 9.4.
\label{TotDminS}
\end{equation}
Accordingly, the average error of the distances computed for these seven satellites is $\simeq 1.3\%$.

In the case of Uranus we implement A[n] to the tetrad: Miranda, Ariel, Titania, and Oberon.  The prescribed ratios are now (using the symbols of Table~\ref{uranus})  
\begin{equation}
R_{1,2} = A_\mathrm{M}/A_\mathrm{A}, \ \ R_{1,3} = A_\mathrm{M}/A_\mathrm{T}, \ \ 
                                         R_{1,4} = A_\mathrm{M}/A_\mathrm{O}.
\label{ratiosU}
\end{equation}
Table~\ref{uranus} gives for the optimum case of Uranus
\begin{equation}
\Delta_\mathrm{min} \biggl( n_\mathrm{opt}(\mathrm{U}) = 1.213; \, q_\mathrm{M}=5, \, 
                    q_\mathrm{A}=7, \, q_\mathrm{T}=14, q_\mathrm{O}=18 \biggr) 
                    \simeq 2.0.
\label{DminU}
\end{equation}
Here, smaller error is that for Oberon's distance, $\simeq 0.1\%$, while larger one is that for Ariel's distance, $\simeq 1.3\%$; note that Oberon is also the most massive satellite of Uranus. The average error in the computed distances of the four satellites is $\simeq 0.5\%$.

Now, concerning the satellite Umbriel, not belonging to the initial tetrad, we find a discrepancy $\simeq 1.8\%$ in its computed distance. So, the total sum of percent errors for the five satellites of Uranus considered here becomes 
\begin{equation}
\Delta_\mathrm{tot}(\mathrm{U}) \simeq 3.8.
\label{TotDminU}
\end{equation}
The average error in the distances computed for these five satellites is $\simeq 0.9\%$.

Finally, Table~\ref{neptune} gives numerical results for the optimum case of Neptune. We first clarify that, in this case, we implement A[n] to the tetrad: Galatea, Proteus, Triton, and Nereid. So, the prescribed ratios are (using the symbols of Table~\ref{neptune})
\begin{equation}
R_{1,2} = A_\mathrm{G}/A_\mathrm{P}, \ \ R_{1,3} = A_\mathrm{G}/A_\mathrm{P}, \ \ 
                                         R_{1,4} = A_\mathrm{G}/A_\mathrm{Ne}.
\label{ratiosN}
\end{equation}
For the Neptune's optimum case we find
\begin{equation}
\Delta_\mathrm{min} 
\biggl( n_\mathrm{opt}(\mathrm{N}) = 1.124; \, q_\mathrm{G}=3, \, 
       q_\mathrm{P}=5, \, q_\mathrm{T}=13, \, q_\mathrm{Ne}=166 \biggr) \simeq 2.7.
\label{DminN}
\end{equation}
The smaller error is that for Nereid's distance, $\simeq 0.04\%$, which is also the most massive satellite of Neptune; while the larger error is that for Galatea's distance, $\simeq 1.3\%$. The average error in the distances of the four satellites is $\simeq 0.7\%$. 

For the satellite Larissa, which does not belong to the initial tetrad, we compute a distance suffering a discrepancy $\simeq 14.6\%$. It is well remarking here that Larissa seems to be accomodated inside the shell No~3, which does also contain the satellite Galatea. This reason seems to contribute to the relatively large error on Larissa's distance. The total sum of percent errors for the five satellites of Neptune considered here becomes 
\begin{equation}
\Delta_\mathrm{tot}(\mathrm{N}) \simeq 17.4.
\label{TotDminN}
\end{equation}
Accordingly, the average error in the computed distances for these five satellites is $\simeq 3.5\%$.

\begin{table}
\begin{center}
\caption{Jupiter's system of satellites: the central body $S_1$, i.e. Jupiter (J), and the polytropic spherical shells of the Galilean satellites Io (I), Europa (E), Ganymede (G), and Callisto (C). All radii, except for $\xi_1$, are measured in planet's radii $\xi_1$. Percent errors in the computed radii $\alpha$ of the satellite orbits are given with respect to the corresponding observed radii $A$, $100 \times |(A - \alpha)| / A$.\label{jupiter}}
\begin{tabular}{lr} 
\hline \hline
Jupiter--Shell No                    & 1                 \\
$n_\mathrm{opt}(\mathrm{J})$                                                    & $8.930(-01)$      \\
$\xi_1$                              & 3.049570(+00)     \\
$R_\mathrm{J}$ (cm)                                               & 6.991100(+09)     \\
\hline

Io--Shell No                                                      & 7              \\
Inner radius, $\xi_6$                                             & 5.596891(+00)     \\
Outer radius, $\xi_7$                                             & 6.465937(+00)     \\
Radius $\alpha_\mathrm{I}$ of $\mathrm{max}|\theta|$ & 6.033668(+00)     \\
Percent error in $\alpha_\mathrm{I}$, given that $A_\mathrm{I}=6.0320\,R_\mathrm{J}$   
& $2.765(-02)$              \\ 
\hline

Europa--Shell No                                                  & 11                 \\
Inner radius, $\xi_{10}$                                          & 9.090419(+00)     \\
Outer radius, $\xi_{11}$                                          & 9.937953(+00)     \\
Radius $\alpha_\mathrm{E}$ of $\mathrm{max}|\theta|$ & 9.509566(+00)     \\
Percent error in $\alpha_\mathrm{E}$, given that $A_\mathrm{E}=9.5984\,R_\mathrm{J}$   
& $9.255(-01)$              \\ 
\hline

Ganymede--Shell No                                           & 18               \\
Inner radius, $\xi_{17}$                                     & 1.502706(+01)     \\
Outer radius, $\xi_{18}$                                     & 1.587780(+01)     \\
Radius $\alpha_\mathrm{G}$ of $\mathrm{max}|\theta|$   & 1.544483(+01)     \\
Percent error in $\alpha_\mathrm{G}$, given that $A_\mathrm{G}=15.3110\,R_\mathrm{J}$     
& $8.741(-01)$     \\ 
\hline

Callisto--Shell No                                                 & 32               \\
Inner radius, $\xi_{31}$                                           & 2.657665(+01)     \\
Outer radius, $\xi_{32}$                                           & 2.740172(+01)     \\
Radius $\alpha_\mathrm{C}$ of $\mathrm{max}|\theta|$         & 2.698743(+01)     \\
Percent error in $\alpha_\mathrm{C}$, given that $A_\mathrm{C}=26.9300\,R_\mathrm{J}$       
& $2.133(-01)$               \\   
\hline

\end{tabular}
\end{center}
\end{table}

\begin{table}
\begin{center}
\caption{Saturn's system of satellites: the central body $S_1$, i.e. Saturn (S), and the polytropic spherical shells of the satellites Enceladus (E), Tethys (Te), Dione (D), Rhea (R), Titan (T), Hyperion (H), and Iapetus (I). Details as in Table~\ref{jupiter}.\label{saturn}}
\begin{tabular}{lr} 
\hline \hline
Saturn--Shell No                                   & 1                 \\
$n_\mathrm{opt}(\mathrm{S})$                       & 1.239(+00)        \\
$\xi_1$                                            & 3.367926(+00)     \\
$R_\mathrm{S}$ (cm)                                               & 5.823200(+09)     \\
\hline

Enceladus--Shell No                                               & 4              \\
Inner radius $\xi_3$                   & 3.293092(+00)     \\
Outer radius $\xi_4$                   & 4.517413(+00)     \\
Radius $\alpha_\mathrm{E}$ of $\mathrm{max}|\theta|$ & 3.905252(+00)     \\
Percent error in $\alpha_\mathrm{E}$, given that $A_\mathrm{E}=4.0862\,R_\mathrm{S}$   
& $4.430(+00)$              \\ 
\hline

Tethys--Shell No                                                  & 5              \\
Inner radius, $\xi_4$                                             & 4.517413(+00)     \\
Outer radius, $\xi_5$                                             & 5.836690(+00)     \\
Radius $\alpha_\mathrm{Te}$ of $\mathrm{max}|\theta|$ & 5.136720(+00)     \\
Percent error in $\alpha_\mathrm{Te}$, given that $A_\mathrm{Te}=5.0594\,R_\mathrm{S}$   
& $1.528(+00)$              \\ 
\hline

Dione--Shell No                                                   & 6                 \\
Inner radius, $\xi_5$                                             & 5.836690(+00)     \\
Outer radius, $\xi_6$                                             & 7.121010(+00)     \\
Radius $\alpha_\mathrm{D}$ of $\mathrm{max}|\theta|$ & 6.502545(+00)     \\
Percent error in $\alpha_\mathrm{D}$, given that $A_\mathrm{D}=6.4809\,R_\mathrm{S}$   
& $3.400(-01)$              \\ 
\hline

Rhea--Shell No                                                    & 8               \\
Inner radius, $\xi_{7}$                                     & 8.468297(+00)     \\
Outer radius, $\xi_{8}$                                     & 9.825013(+00)     \\
Radius $\alpha_\mathrm{R}$ of $\mathrm{max}|\theta|$  & 9.056045(+00)     \\
Percent error in $\alpha_\mathrm{R}$, given that $A_\mathrm{R}=9.0519\,R_\mathrm{S}$     
& $4.579(-02)$     \\ 
\hline

Titan--Shell No                                                  & 16               \\
Inner radius, $\xi_{15}$                                         & 1.992918(+01)     \\
Outer radius, $\xi_{16}$                                         & 2.141319(+01)     \\
Radius $\alpha_\mathrm{T}$ of $\mathrm{max}|\theta|$       & 2.060618(+01)     \\
Percent error in $\alpha_\mathrm{T}$, given that $A_\mathrm{T}=20.984\,R_\mathrm{S}$       
& $1.801(+00)$               \\   
\hline

Hyperion--Shell No                                               & 19               \\
Inner radius, $\xi_{18}$                                         & 2.444746(+01)     \\
Outer radius, $\xi_{19}$                                         & 2.599817(+01)     \\
Radius $\alpha_\mathrm{H}$ of $\mathrm{max}|\theta|$       & 2.520841(+01)     \\
Percent error in $\alpha_\mathrm{H}$, given that $A_\mathrm{H}=25.433\,R_\mathrm{S}$       
& $8.831(-01)$               \\   
\hline

Iapetus--Shell No                                                & 41               \\
Inner radius, $\xi_{40}$                                         & 6.006676(+01)     \\
Outer radius, $\xi_{41}$                                         & 6.180609(+01)     \\
Radius $\alpha_\mathrm{I}$ of $\mathrm{max}|\theta|$       & 6.092771(+01)     \\
Percent error in $\alpha_\mathrm{I}$, given that $A_\mathrm{I}=61.149\,R_\mathrm{S}$       
& $3.619(-01)$               \\   
\hline

\end{tabular}
\end{center}
\end{table}

\begin{table}
\begin{center}
\caption{Uranus's system of satellites: the central body $S_1$, i.e. Uranus (U), and the polytropic spherical shells of the satellites Miranda (M), Ariel (A), Umbriel (Um), Titania (T), and Oberon (O). Details as in Table~\ref{jupiter}.\label{uranus}}
\begin{tabular}{lr} 
\hline \hline
Uranus--Shell No                                   & 1                 \\
$n_\mathrm{opt}(\mathrm{U})$                       & 1.213(+00)        \\
$\xi_1$                                            & 3.341791(+00)     \\
$R_\mathrm{U}$ (cm)                                               & 2.536200(+09)     \\
\hline

Miranda--Shell No                                                 & 5              \\
Inner radius, $\xi_4$                   & 4.467280(+00)     \\
Outer radius, $\xi_5$                   & 5.747010(+00)     \\
Radius $\alpha_\mathrm{M}$ of $\mathrm{max}|\theta|$ & 5.087121(+00)     \\
Percent error in $\alpha_\mathrm{M}$, given that $A_\mathrm{M}=5.1017\,R_\mathrm{U}$   
& $2.858(-01)$              \\ 
\hline

Ariel--Shell No                                                  & 7              \\
Inner radius, $\xi_6$                                            & 6.999681(+00)     \\
Outer radius, $\xi_7$                                            & 8.309665(+00)     \\
Radius $\alpha_\mathrm{A}$ of $\mathrm{max}|\theta|$ & 7.630666(+00)     \\
Percent error in $\alpha_\mathrm{A}$, given that $A_\mathrm{A}=7.5317\,R_\mathrm{U}$   
& $1.314(+00)$              \\ 
\hline

Umbriel--Shell No                                                 & 9                 \\
Inner radius, $\xi_8$                                             & 9.623268(+00)     \\
Outer radius, $\xi_9$                                             & 1.099735(+01)     \\
Radius $\alpha_\mathrm{D}$ of $\mathrm{max}|\theta|$ & 1.031031(+01)     \\
Percent error in $\alpha_\mathrm{Um}$, given that $A_\mathrm{Um}=10.500\,R_\mathrm{U}$   
& $1.807(+00)$              \\ 
\hline

Titania--Shell No                                                 & 14               \\
Inner radius, $\xi_{13}$                                     & 1.652428(+01)     \\
Outer radius, $\xi_{14}$                                     & 1.790659(+01)     \\
Radius $\alpha_\mathrm{T}$ of $\mathrm{max}|\theta|$   & 1.723629(+01)     \\
Percent error in $\alpha_\mathrm{T}$, given that $A_\mathrm{T}=17.188\,R_\mathrm{U}$     
& $2.810(-01)$     \\ 
\hline

Oberon--Shell No                                                  & 18               \\
Inner radius, $\xi_{17}$                                          & 2.224805(+01)     \\
Outer radius, $\xi_{18}$                                          & 2.366986(+01)     \\
Radius $\alpha_\mathrm{O}$ of $\mathrm{max}|\theta|$        & 2.298171(+01)     \\
Percent error in $\alpha_\mathrm{O}$, given that $A_\mathrm{O}=23.008\,R_\mathrm{U}$       
& $1.143(-01)$               \\   
\hline

\end{tabular}
\end{center}
\end{table}

\begin{table}
\begin{center}
\caption{Neptune's system of satellites: the central body $S_1$, i.e. Neptune (N), and the polytropic spherical shells of the satellites Galatea (G), Larissa (L), Proteus (P), Triton (T), and Nereid (Ne). Details as in Table~\ref{jupiter}.\label{neptune}}
\begin{tabular}{lr} 
\hline \hline

Neptune--Shell No                                  & 1                 \\
$n_\mathrm{opt}(\mathrm{N})$                       & 1.124(+00)        \\
$\xi_1$                                            & 3.255217(+00)     \\
$R_\mathrm{N}$ (cm)                                               & 2.462200(+09)     \\
\hline

Galatea--Shell No                                                 & 3              \\
Inner radius, $\xi_2$                   & 2.058961(+00)     \\
Outer radius, $\xi_3$                   & 3.166719(+00)     \\
Radius $\alpha_\mathrm{G}$ of $\mathrm{max}|\theta|$ & 2.549784(+00)     \\
Percent error in $\alpha_\mathrm{G}$, given that $A_\mathrm{G}=2.5162\,R_\mathrm{N}$   
& $1.335(+00)$              \\ 
\hline

Larissa--Shell No                                                & 3              \\
Inner radius, $\xi_2$                                            & 2.058961(+00)  \\
Outer radius, $\xi_3$                                            & 3.166719(+00)  \\
Radius $\alpha_\mathrm{L}$ of $\mathrm{max}|\theta|$       & 2.549784(+00)  \\
Percent error in $\alpha_\mathrm{L}$, given that $A_\mathrm{L}=2.9871\,R_\mathrm{N}$   
& $1.464(+01)$              \\ 
\hline

Proteus--Shell No                                                 & 5                 \\
Inner radius, $\xi_4$                                             & 4.284590(+00)     \\
Outer radius, $\xi_5$                                             & 5.438126(+00)     \\
Radius $\alpha_\mathrm{P}$ of $\mathrm{max}|\theta|$        & 4.823057(+00)     \\
Percent error in $\alpha_\mathrm{P}$, given that $A_\mathrm{P}=4.7781\,R_\mathrm{N}$   
& $9.409(-01)$              \\ 
\hline

Triton--Shell No                                                  & 13               \\
Inner radius, $\xi_{12}$                                     & 1.375240(+01)     \\
Outer radius, $\xi_{13}$                                     & 1.498369(+01)     \\
Radius $\alpha_\mathrm{T}$ of $\mathrm{max}|\theta|$   & 1.434623(+01)     \\
Percent error in $\alpha_\mathrm{T}$, given that $A_\mathrm{T}=14.408\,R_\mathrm{N}$     
& $4.287(-01)$     \\ 
\hline

Nereid--Shell No                                                  & 166           \\
Inner radius, $\xi_{165}$                                         & 2.233442(+02)     \\
Outer radius, $\xi_{166}$                                         & 2.247699(+01)     \\
Radius $\alpha_\mathrm{Ne}$ of $\mathrm{max}|\theta|$       & 2.240404(+01)     \\
Percent error in $\alpha_\mathrm{Ne}$, given that $A_\mathrm{Ne}=223.94\,R_\mathrm{N}$       
& $4.482(-02)$               \\   
\hline

\end{tabular}
\end{center}
\end{table}

\end{document}